\documentclass[twocolumn,showpacs,superscriptaddress,amsmath,amssymb,pra]{revtex4}
\usepackage{graphicx}
\usepackage{bm}
\usepackage{subfigure}
\usepackage{epstopdf}

\begin{document}

\title{Dimensional Effects on the Momentum distribution of Bosonic Trimer States}
\author{F.~F. Bellotti}
\affiliation{Instituto Tecnol\'{o}gico de Aeron\'autica, 12228-900, S\~ao Jos\'e dos Campos, SP, Brazil}
\affiliation{Instituto de Fomento e Coordena\c{c}{\~a}o Industrial, 12228-901, S{\~a}o Jos{\'e} dos Campos, SP, Brazil}
\author{T. Frederico}
\affiliation{Instituto Tecnol\'{o}gico de Aeron\'autica, 12228-900, S\~ao Jos\'e dos Campos, SP, Brazil} 
\author{M.~T. Yamashita}
\affiliation{Instituto de F\'\i sica Te\'orica, UNESP - Univ Estadual Paulista, C.P. 70532-2, CEP 01156-970, S\~ao Paulo, SP, Brazil} 
\author{D.~V. Fedorov}
\author{A.~S. Jensen}
\author{N.~T. Zinner}
\affiliation{Department of Physics and Astronomy, Aarhus University, DK-8000 Aarhus C, Denmark}
\date{\today }

\begin{abstract}
The momentum distribution is a powerful probe of strongly-interacting
systems that are expected to display universal behavior. This is contained
in the contact parameters which relate few- and many-body properties. 
Here we consider a Bose gas in two dimensions and explicitly show that the 
two-body contact parameter is universal and then demonstrate that the 
momentum distribution at next-to-leading order has a logarithmic 
dependence on momentum which is vastly different from the three-dimensional case.
Based on this, we propose a scheme 
for measuring the effective dimensionality of a quantum many-body system by 
exploiting the functional form of the momentum distribution.
\end{abstract}
\maketitle

\section{Introduction}
Strongly interacting quantum systems are ubiquitous in Nature and 
naturally at the forefront of physics research. However, the theoretical 
study of strong interactions can be very difficult since our usual 
and intuitively clear perturbative methods can fail miserably when 
inter-particle forces are strong. A success of cold atomic gas physics
is the ability to create and manipulate strongly interacting
gases in experiment \cite{bloch2008}. One particularly interesting 
aspect is the ability to change the dimensionality by applied optical 
fields. This means that low-dimensional dynamics which is typically 
found in condensed-matter systems of great technological interest 
can be addressed. The field thus provides a testbed for models of 
strongly-coupled dynamics that are used to describe interesting
materials.

A breakthrough in the study of strongly-interacting quantum systems
with short-range interactions 
was the derivation of a set of universal relations that relate the 
two-body correlations to the many-body thermodynamics through the so-called 
contact parameter, $C_2$ \cite{tan2008}. One way to define this quantity
is through the asymptotic behavior of the single-particle momentum 
distribution, $n(\bm k)$, of a few- or many-body system, i.e. via $n(\bm k)\to C_2/k^4$ which is 
the leading order behavior when
the momentum, $\bm k$, goes to infinity (this fact had been already 
derived for a one-dimensional Bose gas with zero-range interactions, 
the Lieb-Liniger system \cite{olshanii2003}). The same $C_2$ also appears
in the total energy of the system and in response functions. 
These relations were 
subsequently confirmed in experiments on two-component Fermi 
gases \cite{stewart2010,kuhnle2010}. They also hold 
for bosonic gases \cite{combescot2009,schakel2010,braaten2011,castin2011,werner2012}
as confirmed by recent experiments \cite{wild2012}. In the case
of two-component fermions, the Pauli principle suppresses correlations
between three particles. In contrast, for bosons 
three-body correlations are very important
and this implies that one
must consider also a three-body contact 
parameter, $C_3$ \cite{braaten2011,castin2011}. It is most simply
defined as the coefficient of the sub-leading large $\bm k$ limiting 
term in $n(\bm k)$, 
but as we will show below, the form of this term is highly sensitive
to dimensionality.

A second avenue that is enjoying great success at the moment, is the 
experimental study of two-dimensional (2D) atomic Fermi 
gases \cite{turlapov2010,frohlich2011,dyke2011,feld2011,sommer2012}. 
Universal contact relations should also hold in this
case \cite{combescot2009,werner2010,valiente2011,langmack2012,kohl2012}.
Interestingly, 
a recent experiment \cite{vogt2012} has found that the monopole
breathing mode is essentially undamped and has no interaction-dependent
shift \cite{olshanii2010,hofmann2012,taylor2012}. This implies a 
scale-invariance in the system \cite{rosch1997} that has also been 
observed in weakly-interacting 2D Bose gases \cite{hung2011}. However, 
this observation is hard to reconcile with the fact that a scale is provided by the 
energy of the two-body dimer which is always bound for attractive 
short-range interaction in 2D. One would naively expect 
modifications of both few- and many-body dynamics in these systems.

From a few-body perspective, the 
special features of 2D systems are manifest in the spectrum of 
three identical bosons with attractive zero-range interactions
(the so-called universal limit), since
no length scale is provided by the two-body potential except
for the one given by the two-body dimer binding energy, $E_2$. 
Here
one finds that there are {\it exactly} two bound states which have
energies $E_3=16.52E_2$ and $E_3=1.270E_2$
\cite{bru79,adh88,nie97,nie01}. This is in sharp
contrast to 3D, where an infinite set of geometrically separated
states appear at the threshold for two-body binding \cite{efi70}.
In realistic systems, this scaling is broken by the finite-range of the 
interaction \cite{braaten2006}, and one obtains a normalization of the
spectrum since the range determines the lowest bound universal bound state 
(there are deeply bound states that have small radii and non-universal
structure which are not of interest here). Typically one parametrizes the 
short-range physics by introducing the three-body
parameter, $\kappa^\textrm{3D}_{*}$, to get the correct three-body 
energy \cite{braaten2006}. However, in 2D
such a procedure is not needed for three particles, 
i.e. there is no need for a $\kappa^\textrm{2D}_{*}$. In the universal limit 
in 2D this 
implies that the three-body energies must be proportional to the dimer energy. 

In this paper, we study identical bosons in 2D 
with attractive short-range interactions and use few-body methods to 
determine $C_2$ and $C_3$. This is achieved by computing the 
momentum distribution for three identical bosons, in particular
its asymptotic behavior for large momenta. We provide both analytical
and numerical evidence that support a {\it universal} tail behavior that is
the same for both ground and excited states. 
This is the first time that $C_3$ has been 
discussed in 2D to the best of our knowledge. 
Moreover, we show that the 
sub-leading term has a novel behavior that is radically 
different in 2D as compared to 3D. Based on this fact, we 
propose to use the momentum 
distribution to measure the effective dimensionality of 
a quantum system in the universal regime. Our study is thus 
a first step in exploring effects of dimensional 
crossover on higher-order correlations in many-body systems. 

\section{Method}
We consider three identical bosons with mass $m$. We use
attractive two-body interactions of zero range and parameterized
by the dimer binding energy, $E_2$. The two-body T-matrix for energy $E$ is 
thus $\tau(E)=(-2\pi \textrm{ln}\sqrt{-E/E_2})^{-1}$ in units where $\hbar=m=1$ 
\cite{adh88,bel11,bel12}. By using Faddeev decomposition and bosonic symmetry, the 
three-body wave function, $\Psi$, can be written
\begin{align}
\Psi\left(\mathbf{q},\mathbf{p}\right)= \frac{f(q)+f\left( \left| \mathbf{p}- \frac{\mathbf{q}}{2}\right| \right)+f\left( \left| \mathbf{p}+ \frac{\mathbf{q}}{2}\right| \right)}{E_{3}+\mathbf{p}^2+ \frac{3}{4}\mathbf{q}^2},
\label{eq.04}
\end{align}
where $\bm p=\tfrac{1}{2}(\bm k_1-\bm k_2)$ and 
$\bm q=\tfrac{2}{3}\bm k_3-\tfrac{1}{3}(\bm k_1+\bm k_2)$ are Jacobi momenta, $\bm k_i\,,\,i=1,2,3$ the lab momenta, and $E_3$ is the three-body energy. The 
spectator functions, $f(\bm q)$, satisfy the set of integral equations 
\begin{align}
f\left( \mathbf{q}\right) = 2\tau\left(-E_3-\frac{3}{4}\mathbf{q}^{2}\right)\int{d^2k\frac{f\left( \mathbf{k}\right) }{-E_{3}-\mathbf{q}^{2}-\mathbf{k}^{2}-\mathbf{k}\cdot\mathbf{q}}}.
\label{eq.03}
\end{align}
Armed with the solution to this equation, the momentum distribution is 
\begin{align}
n(q)=\int{d^2 p \left| \frac{f(q)+f\left( \left| \mathbf{p}- \frac{\mathbf{q}}{2}\right| \right)+f\left( \left| \mathbf{p}+ \frac{\mathbf{q}}{2}\right| \right)}{E_{3}+\mathbf{p}^2+ \frac{3}{4}\mathbf{q}^2}\right|^2}.
\label{eq.01}
\end{align}
Following the discussion in Ref.~\cite{castin2011}, we define four components
$n^m(q)=\sum_{i=1}^{4}n_{i}^{m}(q)$, where $m=0$ denotes the ground state and 
$m=1$ the excited state. The individual components are
\begin{equation}
n_1^m(q)=f^{2}_{m}(q)\int{d^2p\frac{1}{\left(E_{3}^{m}+\mathbf{p}^2+ \frac{3}{4}\mathbf{q}^2\right)^2}}=\frac{\pi f^{2}_{m}(q)}{E_{3}^{m}+\frac{3}{4}\mathbf{q}^2},
\label{eq.06}
\end{equation}
\begin{eqnarray}
n_2^m(q)
=4f_m(q)\int{d^2k\frac{f_m(k)}{\left(E_{3}^{m}+\mathbf{k}^2+ \mathbf{q}^2+ \mathbf{k}\cdot\mathbf{q}\right)^2}},
\label{eq.07}
\end{eqnarray}
\begin{equation}
n_3^m(q)
=2\int{d^2k\frac{f^{2}_{m}(k)}{\left(E_{3}^{m}+\mathbf{k}^2+ \mathbf{q}^2+ \mathbf{k}\cdot\mathbf{q}\right)^2}},
\label{eq.08}
\end{equation}
\begin{eqnarray}
n_4^m(q)
=2\int{d^2k\frac{f_m(k)f_m\left( \left| \mathbf{k}- \mathbf{q}\right| \right)}{\left(E_{3}^{m}+\mathbf{k}^2+ \mathbf{q}^2+ \mathbf{k}\cdot\mathbf{q}\right)^2}},
\label{eq.09}
\end{eqnarray}
where $m$ on $f(q)$ and $E_3$ labels the state. Throughout, we measure all momenta in units of $\sqrt{E_2}$.
Note that the normalization we use is $\int d^2k n(k)=1$.

\section{Large-momentum limit}
The leading order (LO) behavior of the momentum distribution exhibits the same 
$C_2k^{-4}$ tail in 1D, 2D, and 3D since it derives solely from 
two-body physics \cite{valiente2012}. However, $C_2$ depends on
what system is addressed and whether few-body bound states are present. 
For bosons in 3D, the tail is \cite{castin2011,braaten2011}
\begin{align}
n_{3D}(k)\to \frac{1}{k^4}C_2+\frac{\cos[2s_0 \textrm{ln}(\sqrt{3}k/\kappa_*)+\phi]}{k^5}C_3,
\label{n3d}
\end{align}
where $s_0=1.00624$ and $\phi=-0.87280$ are constants that can be 
determined from a full solution of the three-bosons problem in 3D at 
unitarity \cite{castin2011} with trimer energy $E_3=\kappa^{2}_{*}$
(using $\kappa_{*}^\textrm{3D}=\kappa_*$ for simplicity). The log-periodic
three-body next-to-leading order (NLO) term derives from the Efimov effect, whose
solution can be used to determine $3(2\pi)^3 C_2=53.097/\kappa_*$ and $3(2\pi)^3 C_3=-89.263/\kappa_{*}^{2}$
\cite{castin2011}. The factor $3(2\pi)^3$ is due to a difference in  
definition of $n(k)$ in Eq.~\eqref{eq.01} in comparison to Ref.~\cite{castin2011}.
As discussed above, in 2D there is no Efimov effect for three bosons. The 
log-periodic behavior is therefore not expected {\it a priory}. As we will 
now demonstrate, the distribution in 2D is very different. It has
the structure
\begin{align}
n_{2D}(k)\to \frac{1}{k^4}C_2+\frac{\textrm{ln}^3(k)}{k^6}C_3,
\label{n2d}
\end{align}
and we see indeed a very different NLO term. We note that the 
NLO term is different from the fermionic case discussed in Ref.~\cite{werner2010} where
no $\textrm{ln}(k)$ factors are present and implies that quantum statistics plays
a role in determining the functional form of the NLO term.
Furthermore, it implies that the NLO term
is in fact an effective measure of dimensionality of bosonic systems in the 
universal regime. We will return to this point below. 

To derive the tail behavior in  Eq.~\eqref{n2d}, one needs to determine
first the spectator function, $f_m(\bm q)$, in Eq.~\eqref{eq.03} for large 
$q$. This can be done analytically and we provide the details 
in Appendix~\ref{appA}. The result is that $f_m(q)\to A_m\textrm{ln}(q)/q^2$,
where $A_m$ is a state-dependent constant. 
This function can now be inserted into Eqs.~\eqref{eq.06} to \eqref{eq.09} 
and the momentum tail can be determined.
The technical details 
are given in Appendix~\ref{appB}. After the dust settles, the tail 
behaviors can be written
\begin{align}
&n_{1}^{m}(q)\to \frac{4\pi}{3}\frac{A_{m}^{2}\textrm{ln}^2(q)}{q^6}\, ,\, 
n_{2}^{m}(q)\to 4\pi\frac{A_{m}^{2}\textrm{ln}^3(q)}{q^6}&\nonumber \\
&n_{3}^{m}(q)\to \frac{4\pi}{q^4}\int_{0}^{\infty}dk k f^{2}_{m}(k)\, ,\, 
n_{4}^{m}(q)\to 2\pi\frac{A_{m}^{2}\textrm{ln}^3(q)}{q^6}.&\nonumber
\end{align}
The LO term clearly comes from $n_{3}^{m}(q)$, while NLO has 
contributions from $n_{2}^{m}(q)$ and $n_{4}^{m}(q)$. However,
there is an additional complication as NLO will also come from
$n_{3}^{m}$ at the next order (not shown above). More 
precisely, we need to determine 
\begin{align}
D_m=\lim_{q\to\infty}\left[n_{3}^{m}-\frac{C_2}{q^4}\right]\frac{q^6}{\textrm{ln}^3(q)},
\end{align}
where $C_2=\lim_{q\to\infty}q^4n^{m}(q)$ which is independent of $m$ as
we discuss below.
We find that $D_m$ is a non-zero constant that depends on the 
state $m$, which means that $C_3$ in 2D should be denoted $C_{3}^{m}$. 
Adding the NLO contributions from $n_{2}^{m}(q)$, $n_{3}^{m}(q)$
and $n_{4}^{m}(q)$, we find
\begin{align}
C_3^{0}=52.07\,\textrm{and}\, C_{3}^{1}=1.01
\end{align}
This $m$-dependence is absent in 3D for $\kappa_*\to 0$ 
but at the cost of the log-periodic
term due to the Efimov effect \cite{castin2011}. That the present 2D case
has state-dependence is a result of the lack of geometric scaling symmetry
in 2D. Note that the next order comes from $n_{1}^{m}(q)$ and 
differs by one power of $\textrm{ln}(q)$ compared to the NLO term.

\begin{figure}[ht!]
\includegraphics[scale=0.29]{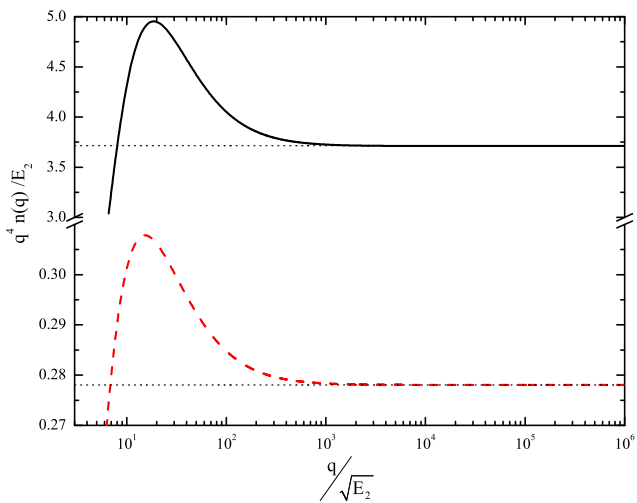}
\caption{LO momentum distribution tail, $q^4 n(q)$, for ground 
(upper solid black line) and excited (lower dashed red line) 
three-body states. Note that the vertical axis is not uniform. 
The asymptotic dashed lines are the analytical results discussed in 
Appendix~\ref{appB}.}
\label{fig1}
\end{figure}

\subsection{Universal behavior}
The LO behavior in 2D is characterized by $C_2$. Explicitly, we have
\begin{align}
n_{3}^{0}(q)\to \frac{3.71E_2}{q^4}\,\textrm{and}\, n_{3}^{1}(q)\to \frac{0.28E_2}{q^4}.
\label{LOtail}
\end{align}
These results have been obtained analytically (see Appendix~\ref{appB}). We have
also done a numerical check which is shown in Fig.~\ref{fig1}. The 
units in Eq.~\eqref{LOtail} are, however, not natural in the same way that
is seen in Eq.~\eqref{n3d} where $\kappa_*$ provides the overall scale. 
The natural scale is $E_3$, and using this we find $3.71E_2/16.52E_2=0.224$
and $0.28E_2/1.270E_2=0.219$ for ground and excited state respectively. 
This is a striking results that demonstrates the state-independence of 
the LO term in 2D to within our numerical accuracy of about 2\%. 
We thus predict that the two-body contact for a bosonic system in 2D with 
short-range attractive interactions is 
\begin{align}
C_2/E_3=0.222\pm 0.003,
\end{align}
where $E_3$ is the trimer energy. This should be compared
to the relation $\tfrac{dE}{d\textrm{ln}a}=\pi N C_2$
derived on general grounds in Ref.~\cite{werner2010}. Here the 
factor $N$ appears due to our normalization which is different 
from Ref.~\cite{werner2010}. We find agreement with this result
within our numerical accuracy.

The universal tail behavior is far from trivial. 
In 3D and at unitarity, the discrete
scale invariance induced by the divergence of the three-body problem, 
implies that the system should behave similarly irrespective of which 
trimer state one considers.
This 
does not occur in 2D and the universal trimer energies are in some sense 
magic numbers multiplying the only scale available, $E_2$. Our
results show that in spite of this major difference, the 2D momentum 
tail displays universal behavior, i.e. $C_2/E_3$ has the same 
value for both ground and excited states.

\section{Dimensional crossover}
Comparing the expressions in Eq.~\eqref{n3d} and Eq.~\eqref{n2d}, 
we see the same LO behavior at large momenta, but a 
vastly different NLO term. The oscillations seen in Eq.~\eqref{n3d}
can be traced directly to the discrete scaling symmetry, or more
precisely, the breakdown of scale-invariance in the system. It is 
known that the condition on the dimension, $D$, for this behavior
is $2.3<D<3.8$ \cite{nie97,nie01}. If we imagine an interpolation
between 2D and 3D, we would expect to see log-periodic terms
in this range of $D$. The NLO term is therefore a 
tell-tale sign of effective dimensionality of the system as we 
will now discuss.

\begin{figure}[ht!]
\includegraphics[scale=0.29]{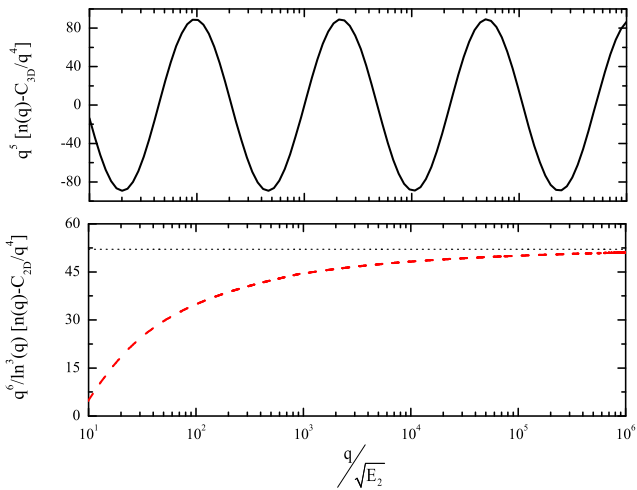}
\caption{NLO momentum distribution comparison of 3D (upper panel) and 
2D (lower panel). The 2D momentum distribution
is the one of the ground state, but the result is similar for the 
excited state.}
\label{fig2}
\end{figure}

In experiments that study cold 2D quantum gases, one uses a tight
transverse optical lattice potential to reduce the motion in this
direction \cite{bloch2008}. As recent experiments have beautifully
demonstrated, the strength of the transverse optical lattice 
can be used to interpolate between 2D and 3D behavior of 
fermionic two-component systems \cite{dyke2011,sommer2012}. Here
we are concerned with bosonic systems, and our results above 
demonstrate how one can use the tail and in particular the NLO 
part of the momentum distribution as a measure of the effective
dimensionality felt by the particles in the system by identifying 
the presence of log-periodic behavior. In Fig.~\ref{fig2} 
we show the extreme cases of 2D and 3D where the 
log-periodic oscillations are clearly seen in the latter, 
while the former has a smooth behavior.

A measurement of the overall functional form of the NLO term is
thus enough to determine the effective dimensionality of the 
squeezed bosonic gas. In a real experiment, the motion in the 
transverse dimension is of course quantized by the lattice, 
and to get a full quantitative understanding this must be taken 
into account (see for instance Ref.~\cite{baur2012}). 
However, since experiments have shown that it is 
possible to reach both the extreme 2D and the 3D regime, there must
necessarily be a dimensional crossover that can be seen in the 
NLO behavior. Of course, from a theoretical point of view it would 
be very attractive to be able to map the strength of the transverse
confinement into some effective dimensionality $D_\textrm{eff}$ which
could be non-integer \cite{valiente2012}.

\section{Experimental implementation}
As we have demonstrated, the NLO term in the momentum 
distribution carries a tell-tale signature of the dimensionality of 
the quantum system under study. The 2D to 3D crossover 
is of immense interest at the moment \cite{dyke2011,sommer2012,baur2012}, 
and it has been shown that both the 3D and the 
strict 2D limits are accessible in experiment. Here we have
discussed the crossover by using formalism applicable to 
either pure 2D or pure 3D without explicit consideration of the
external confinement. Our results predict that a proof-of-principle
experiment is possible by going to the two strict limits. However, 
the full crossover including the intermediate regime (quasi-2D)
where the transverse confinement must be taken explicitly into
account is experimentally addressable and should be 
explored theoretically in the future.

To connect our results directly to current experiments, we need
to consider our units, the dimer binding energy $E_2$, and the 
effects of the transverse confinement on this two-body bound state.
The interaction is controlled by Feshbach resonances \cite{chin2010}. 
However, under the confinement, the dimer energy is modified and 
becomes $E_2=B\hbar\omega_z\exp(-\sqrt{2\pi}l_z/|a|)/\pi$ \cite{petrov2001,pricoupenko2007}.
Here $\omega_z$ is the transverse harmonic confinement frequency, $l_z=\sqrt{\hbar/m\omega_z}$ 
the trapping length, $a$ the 3D scattering length associated with the 
Feshbach resonance, and $B=0.905$ is a constant. This formula holds for $a<0$
and $|a|\ll l_z$, while on resonance, $|a|\to\infty$, $E_2=0.244\hbar\omega_z$.
Corrections arise from the non-harmonic optical lattice \cite{orso2005}, 
but they are not essential for our purposes. 
The dimer energy scale can be 
converted into a momentum scale, $k_0$, 
through $E_2=\hbar^2k_{0}^{2}/2m$. To access the tail behavior and
the 2D-3D crossover, we see from Figs.~\ref{fig1} and \ref{fig2}
that the range $k\sim 10^1-10^3 k_0$ is sufficient. Recent 2D 
Bose gas experiments \cite{hung2011,yefsah2011}, use $l_z\sim 3800a_0$,
where $a_0$ is the Bohr radius, which implies that $k_0\sim 10^{-4}a_{0}^{-1}$
when $|a|=\infty$. For the momentum distribution measurements \cite{stewart2010,kuhnle2010},
the maximum momentum reported is about $k\sim 10^{-3}a_{0}^{-1}$. This 
implies that an order of magnitude or two beyond the reported capabilities
is necessary. However, if $a$ is tuned away from resonance to the $a<0$ side, 
$E_2$ will decrease rapidly according to the formulas above, inducing
a corresponding rapid decrease of $k_0$ which should render the 
physics discussed here within reach of current experimental setups.  
Notice that the van der Waals length scale of about 100$a_0$ is 
in the deep tail, so there is no conflict with the universal 
zero-range description employed here.

\section{Summary and Outlook}
We have taken a first step in the study of higher-order correlations
and dimensional crossover by demonstrating how trimer observables in
strongly-interacting quantum gases can be used to probe dimensionality.
Specifically, we see the breakdown of scale-invariance directly in 
the functional form of the tail of the momentum distribution.

A clear direction for future study is a full inclusion of the transverse
direction and the discrete spectrum it brings. We have shown that 
a crossover with fundamental influence on the momentum tail will 
happen, but mapping it out in a system that is squeezed by optical
lattice(s) is the next task. This would also be interesting for the 
1D-3D or 1D-2D crossovers. Another venue to explore is mass imbalanced
systems where the spectrum is known to be more rich in 2D than the 
equal mass case \cite{bel11,bel12}.

\acknowledgments This work was partly support by funds
provided by FAPESP (Funda\c c\~ao de Amparo \`a Pesquisa do Estado
de S\~ao Paulo) and CNPq (Conselho Nacional de Desenvolvimento
Cient\'\i fico e Tecnol\'ogico ) of Brazil, and by the Danish 
Agency for Science, Technology, and Innovation.

\appendix

\section{Asymptotic form of $f(q)$}\label{appA}
Here we give the technical details of the analytical and numerical determination of the three-body wave function and momentum distributions. We will use units $\hbar=m=1$ and all energies are given in units of the two-body dimer energy $E_2$ implying that all momenta are in units of $\sqrt{E_2}$. 
In the symmetric case, where the three masses and the three two-body binding energies are set equal to one, 
the spectator function fulfills the integral equation
\begin{equation}
f(q)=\frac{1/ \pi}{\ln \left(\sqrt{E_3+3/4q^2}\right)}\int{d^2k\frac{f(k)}{E_3+q^2+k^2+\mathbf{k}\cdot\mathbf{q}}}.
\label{eq.A01}
\end{equation}
This can be cast into the useful form
\begin{align}
&f(q)=\frac{\alpha(q,E_3)}{\pi}&\nonumber\\
&\times\int_0^\infty{dk \frac{k f(k)}{E_3+q^2+k^2}}\int_0^{2\pi}{\frac{d\theta}{1+a\cos\theta}},&
\label{eq.A02}
\end{align}
with $a=kq/(E_3+q^2+k^2)$ and where we have defined
\begin{align}
\alpha(q,E_3)=\frac{1}{\ln \left(\sqrt{E_3+3/4q^2}\right)}
\end{align}
The angular integral is
\begin{equation}
\int_0^{2\pi}{\frac{d\theta}{1+a\cos\theta}}=\frac{2\pi}{\sqrt{1-a^2}} \ \ \ \text{for} \ \ \ 0<a<1
\label{eq.A03}
\end{equation}
and one obtains
\begin{align}
&f(q)=2\alpha(q,E_3)&\nonumber\\
&\times\int_0^\infty{dk \frac{k f(k)}{(E_3+q^2+k^2)\sqrt{1-\frac{q^2k^2}{(E_3+q^2+k^2)^2}}}},&
\label{eq.A04}
\end{align}
which can be rewritten as
\begin{align}
f(q)=&2\alpha(q,E_3) \left[\int_0^\Lambda{dk \frac{k f(k)}{(E_3+q^2+k^2)\sqrt{1-\frac{q^2k^2}{(E_3+q^2+k^2)^2}}}} \right.& \nonumber\\
&\left.+\int_\Lambda^\infty{dk \frac{k f(k)}{(E_3+q^2+k^2)\sqrt{1-\frac{q^2k^2}{(E_3+q^2+k^2)^2}}}}\right],&
\label{eq.A05}
\end{align}
where $\Lambda$ is a large-momentum cut-off that will be useful below.
Taking $\Lambda>>\sqrt{E_3}$, the spectator function in Eq.~\eqref{eq.A05} is approximately given by
\begin{align}
f(q)\approx&2\alpha(q,E_3)\int_0^\Lambda{dk \frac{k f(k)}{(E_3+q^2+k^2)\sqrt{1-\frac{q^2k^2}{(E_3+q^2+k^2)^2}}}}& \nonumber\\
+&2\alpha(q,E_3) \int_\Lambda^\infty{dk \frac{k f(k)}{(q^2+k^2)\sqrt{1-\frac{q^2k^2}{(q^2+k^2)^2}}}}.&
\label{eq.A06}
\end{align}
For $q\rightarrow\infty$, the first term on the right-hand-side of Eq.~\eqref{eq.A06} tends to zero in the 
following manner
\begin{equation}
f_1(q)\approx\frac{2}{q^2 \ln q}\int_0^\Lambda{dk \frac{k f(k)}{\sqrt{1-\frac{q^2k^2}{(q^2+k^2)^2}}}}.
\label{eq.A07}
\end{equation}

Now, we assume that
\begin{equation}
f(q)\rightarrow_{q\rightarrow\infty} \frac{\ln q}{q^2} \ .
\label{eq.A08}
\end{equation}
Inserting this ansatz and taking the limit $q\rightarrow\infty$ in the second term on the right-hand-side of Eq.~\eqref{eq.A06} one finds
\begin{align}
&f_2(q)\approx\frac{2}{\ln q}\int_\Lambda^\infty{dk \frac{k \ln k}{k^2 (q^2+k^2) \sqrt{1-\frac{q^2k^2}{(q^2+k^2)^2}}}}\to&\nonumber\\
&\frac{2}{\ln q}\int_\Lambda^\infty{dk \frac{\ln k}{k (q^2+k^2)}},&
\label{eq.A09}
\end{align}
when $q\to \infty$.
Changing variables to $y=k/q$, the second spectator function term becomes
\begin{equation}
f_2(q)\approx\frac{2}{q^2 \ln q}\int_{\Lambda/q}^\infty{\frac{dy}{y} \frac{\ln y + \ln q}{(1+y^2)}},
\label{eq.A11}
\end{equation}
which can be split into
\begin{align}
f_2(q)\approx&\frac{2}{q^2 \ln q}\left[\int_{\Lambda/q}^\infty{\frac{dy}{y} \frac{\ln y}{(1+y^2)}}\right.&\nonumber\\
&\left.+\ln q \int_{\Lambda/q}^\infty{\frac{dy}{y} \frac{1}{(1+y^2)}}\right].&
\label{eq.A12}
\end{align} 
The first integral term on the right-hand-side of Eq.~\eqref{eq.A12} is
\begin{align}
\int_{\Lambda/q}^\infty{\frac{dy}{y} \frac{\ln y}{(1+y^2)}}=&\left. \frac{1}{2} \frac{\ln^2 y}{1+y^2}\right|_{\Lambda/q}^\infty+\int_{\Lambda/q}^\infty{dy \frac{y \ln^2 y}{(1+y^2)^2}}\to &\nonumber\\
&-\frac{1}{2}\ln^2 \frac{\Lambda}{q}=-\frac{1}{2}\ln^2 q&
\label{eq.A13}
\end{align}
for $q\to\infty$. The second term on the right-hand-side of Eq.(\ref{eq.A12}) is
\begin{align}
\int_{\Lambda/q}^\infty{\frac{dy}{y} \frac{1}{(1+y^2)}}=&\left. \frac{\ln y}{1+y^2}\right|_{\Lambda/q}^\infty+2\int_{\Lambda/q}^\infty{dy \frac{y \ln y}{(1+y^2)^2}}&\nonumber\\
&\to -\ln \frac{\Lambda}{q}=\ln q&
\label{eq.A14}
\end{align}
for $q\to\infty$.
Inserting the results of Eqs.~\eqref{eq.A13}) and \eqref{eq.A14} in Eq.~\eqref{eq.A11}) we arrive at
\begin{equation}
f_2(q)\approx\frac{2}{q^2 \ln q}(\ln^2 q-\frac{1}{2}\ln^2 q)=\frac{\ln q}{q^2}.
\label{eq.A15}
\end{equation}
Collecting the results Eqs.~\eqref{eq.A07} and \eqref{eq.A15} we conclude that the ansatz in Eq.~\eqref{eq.A08} give us the asymptotic behavior of the exact spectator function.

\begin{figure}[htb!]
\centering
\includegraphics[width=0.48\textwidth]{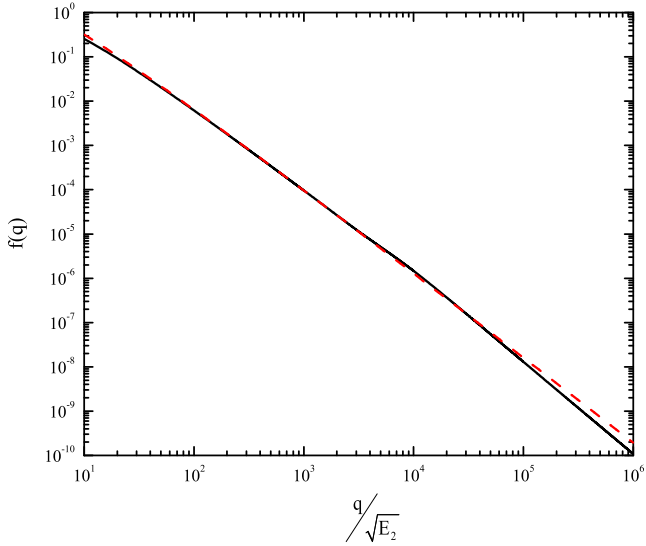}
\caption{Spectator function, $f(q)$, for the ground state calculated numerically (black solid line) and using the ansatz  $f(q)=A_0\frac{\ln q}{q^2}$ (red dashed line).
The solid (black) line tends to oscillates around the dashed (red) one as $q\rightarrow\infty$ due to finite numerical precision.}
\label{fig.A01}
\end{figure}

In Fig.~\ref{fig.A01} both spectator functions obtained from the numerical solution of the set of coupled integral equations and the spectator function asymptotic behavior given in Eq.(\ref{eq.A08}) for the ground state are shown. The log-log scale show us that both magnitude and line's inclination are very close in the region of $q$ between $100$ and $2000$ for the numerical and analytical calculations. For $q>2000$, the curve which represents the numerical solution of the integral equations starts to oscillate around the analytical form.

\section{Momentum density asymptotic behavior}\label{appB}
The one-body momentum distribution is given by
\begin{align}
n(q)=\int{d^2 p \left| \frac{f(q)+f\left( \left| \mathbf{p}- \frac{\mathbf{q}}{2}\right| \right)+f\left( \left| \mathbf{p}+ \frac{\mathbf{q}}{2}\right| \right)}{E_{3}+\mathbf{p}^2+ \frac{3}{4}\mathbf{q}^2}\right|^2},
\end{align}
and can be split into four parts through
\begin{equation}
n^m(q)=\sum_{l=1}^{4}{n_l^m(q)},
\label{eq.B05}
\end{equation}
where the subscript $m$ distinguish ground ($m=0$) and excited states ($m>0$). The individual expressions are 
\begin{equation}
n_1^m(q)=(f_m(q))^2\int{d^2p\frac{1}{\left(E_{3}^{m}+\mathbf{p}^2+ \frac{3}{4}\mathbf{q}^2\right)^2}}=\frac{\pi (f_m(q))^2}{E_{3}^{m}+\frac{3}{4}\mathbf{q}^2},
\label{eq.B06}
\end{equation}
\begin{eqnarray}
n_2^m(q)=&4f_m(q) \int{d^2p \frac{f_m\left( \left| \mathbf{p}+ \frac{\mathbf{q}}{2}\right| \right)}{\left(E_{3}^{m}+\mathbf{p}^2+ \frac{3}{4}\mathbf{q}^2\right)^2}}&\nonumber\\
=&4f_m(q)\int{d^2k\frac{f_m(k)}{\left(E_{3}^{m}+\mathbf{k}^2+ \mathbf{q}^2+ \mathbf{k}\cdot\mathbf{q}\right)^2}}&,
\label{eq.B07}
\end{eqnarray}
\begin{align}
&n_3^m(q)=2\int{d^2p\frac{(f_m\left( \left| \mathbf{p}+ 
\frac{\mathbf{q}}{2}\right| \right))^2}{\left(E_{3}^{m}+\mathbf{p}^2+ \frac{3}{4}\mathbf{q}^2\right)^2}}&\nonumber\\
&=2\int{d^2k\frac{(f_m(k))^2}{\left(E_{3}^{m}+\mathbf{k}^2+ \mathbf{q}^2+ \mathbf{k}\cdot\mathbf{q}\right)^2}},&
\label{eq.B08}
\end{align}
\begin{align}
n_4^m(q)=&2\int{d^2p\frac{f_m\left( \left| \mathbf{p}+ \frac{\mathbf{q}}{2}\right| \right)f_m\left( \left| \mathbf{p}- \frac{\mathbf{q}}{2}\right| \right)}{\left(E_{3}^{m}+\mathbf{p}^2+ \frac{3}{4}\mathbf{q}^2\right)^2}}&\nonumber\\
=&2\int{d^2k\frac{f_m(k)f_m\left( \left| \mathbf{k}- \mathbf{q}\right| \right)}{\left(E_{3}^{m}+\mathbf{k}^2+ \mathbf{q}^2+ \mathbf{k}\cdot\mathbf{q}\right)^2}}&.
\label{eq.B09}
\end{align}
Here we are interested in the limit $q\rightarrow\infty$, where we find
\begin{equation}
n_1^m(q)\approx\frac{4 \pi}{3} \frac{(f_m(q))^2}{q^2},
\label{eq.B10}
\end{equation}
\begin{equation}
n_2^m(q)\approx \frac{4\pi}{q^2} (f_m(q))^2\ln \left(\sqrt{E_{3}^{m}+\frac{3}{4}\mathbf{q}^2}\right)\approx \frac{4\pi}{q^2} (f_m(q))^2\ln(q),
\label{eq.B11}
\end{equation}
\begin{equation}
n_3^m(q)\approx \frac{4 \pi}{q^4}\int_0^\infty{dk k (f_m(k))^2},
\label{eq.B12}
\end{equation}
\begin{equation}
n_4^m(q)\approx \frac{2\pi}{q^2} (f_m(q))^2\ln \left(\sqrt{E_{3}^{m}+\frac{3}{4}\mathbf{q}^2}\right)\approx \frac{2\pi}{q^2} (f_m(q))^2\ln(q).
\label{eq.B13}
\end{equation}
Here the subscript on $f_m(q)$ and the superscript on $E_{3}^{m}$ indicate the state under consideration ($m=0$ for ground and $m=1$ for 
excited state).

The asymptotic form of $f(q)$ when $q\rightarrow\infty$ is (see derivation in Appendix~\ref{appA}) 
\begin{equation}
f_m(q)\rightarrow_{q\rightarrow\infty}A_m \ \frac{\ln q}{q^2}.
\label{eq.B14}
\end{equation}
The functional form of asymptotic behavior is the same for both ground and excited states. However, the normalization 
constant,
\begin{equation}
A_m=\lim_{q\rightarrow\infty}f_m(q)\frac{q^2}{\ln q},
\label{eq.B25}
\end{equation}
is different. Since we are using the normalization $\int d^2k  n(k)=1$, one finds
\begin{equation}
A_0 \approx 1.800 \ \ \ \text{and}  \ \ \ A_1 \approx 0.251 \ , 
\label{eq.B20}
\end{equation}
The numerical results for the asymptotics are shown in Fig.~\ref{fig.B02}.

\begin{figure}[htb!]
\centering
\includegraphics[width=0.48\textwidth]{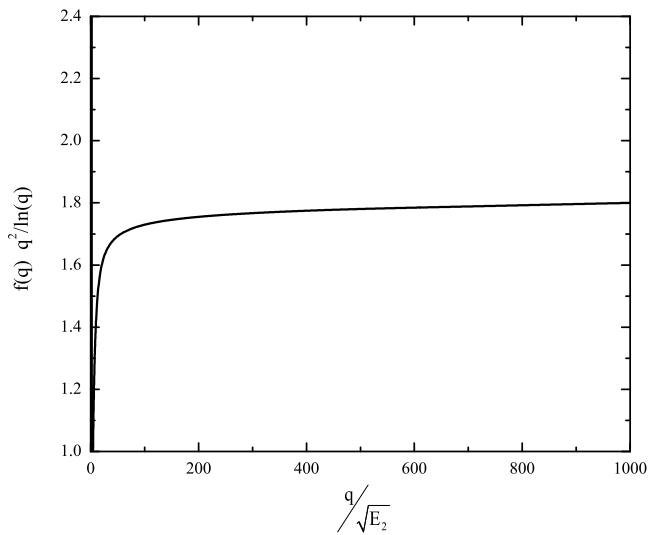}
\caption{Asymptotic behavior of $f(q)$ for the ground state.}
\label{fig.B01}
\end{figure}

Inserting Eq.~\eqref{eq.B14}) in Eqs.~\eqref{eq.B10})-\eqref{eq.B13} one obtains the normalized asymptotic behavior 
\begin{align}
&n_1^m(q)\rightarrow A_m^2 \ \frac{4 \pi}{3} \frac{\ln^2(q)}{q^6},
\label{eq.B15}&\\
&n_2^m(q)\rightarrow A_m^2 \  4\pi \frac{\ln^3(q)}{q^6},
\label{eq.B16}&\\
&n_3^m(q)\rightarrow \frac{C_m}{q^4} \ \text{with} \ C_m=\int_0^\infty{dk k (f_m(k))^2},
\label{eq.B17}&\\
&n_4^m(q)\rightarrow A_m^2 \ 2\pi \frac{\ln^3(q)}{q^6}.
\label{eq.B18}&
\end{align}
The normalization constants given in Eq.~\eqref{eq.B20} determine the asymptotic values and behaviors of the partial momentum density in Eqs.(\ref{eq.B15})-(\ref{eq.B18}) when $q\rightarrow\infty$.

\begin{figure}[htb!]
\centering
\subfigure{\includegraphics[width=0.48\textwidth]{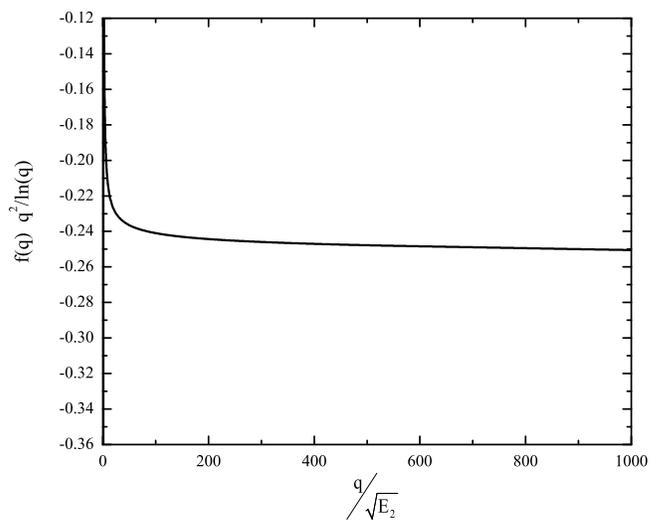}}
\caption{Same as Fig.~\ref{fig.B01} for the first excited state.}
\label{fig.B02}
\end{figure}

\end{document}